
%
%
%
%
%
%

\message{Please respond `b' to query }

%
\input harvmac.tex
%
\global\newcount\mthno \global\mthno=1
\global\newcount\mexno \global\mexno=1
\global\newcount\mquno \global\mquno=1
\def\newsec#1{\global\advance\secno by1\message{(\the\secno. #1)}
\global\subsecno=0\xdef\secsym{\the\secno.}\global\meqno=1\global\mthno=1
\global\mexno=1\global\mquno=1
\bigbreak\medskip\noindent{\bf\the\secno. #1}\writetoca{{\secsym} {#1}}
\par\nobreak\medskip\nobreak}
\xdef\secsym{}
\global\newcount\subsecno \global\subsecno=0
\def\subsec#1{\global\advance\subsecno by1\message{(\secsym\the\subsecno. #1)}
\bigbreak\noindent{\it\secsym\the\subsecno. #1}\writetoca{\string\quad
{\secsym\the\subsecno.} {#1}}\par\nobreak\medskip\nobreak}
\def\appendix#1#2{\global\meqno=1\global\mthno=1\global\mexno=1%
\global\mquno=1
\global\subsecno=0
\xdef\secsym{\hbox{#1.}}
\bigbreak\bigskip\noindent{\bf Appendix #1. #2}\message{(#1. #2)}
\writetoca{Appendix {#1.} {#2}}\par\nobreak\medskip\nobreak}
%
%
\def\thm#1{\xdef #1{\secsym\the\mthno}\writedef{#1\leftbracket#1}%
\global\advance\mthno by1\wrlabeL#1}
\def\que#1{\xdef #1{\secsym\the\mquno}\writedef{#1\leftbracket#1}%
\global\advance\mquno by1\wrlabeL#1}
\def\exm#1{\xdef #1{\secsym\the\mexno}\writedef{#1\leftbracket#1}%
\global\advance\mexno by1\wrlabeL#1}
%
%
\def\ref{\the\refno\nref}
\def\nref#1{\xdef#1{\the\refno}\writedef{#1\leftbracket#1}%
\ifnum\refno=1\immediate\openout\rfile=refs.tmp\fi
\global\advance\refno by1\chardef\wfile=\rfile\immediate
\write\rfile{\noexpand\item{[#1]\ }\reflabeL{#1\hskip.31in}\pctsign}\findarg}
\def\bref{\nref}
%

%
%
%

\def\p{\partial}
%
%
\def\al{\alpha} \def\be{\beta} \def\ga{\gamma}  
\def\de{\delta}  \def\De{\Delta} 
   \def\th{\theta}
    \def\ka{\kappa}
\def\la{\lambda}  \def\rh{\rho} 
    
\def\ph{\phi}     \def\ch{\chi}
\def\ps{\psi}  \def\Ps{\Psi}   
%
%

%
 
%
%
  \def\cF{{\cal F}} 
   
  \def\cW{{\cal W}} 
%
%
%
\def\lefthook{{\vrule height5pt width0.4pt depth0pt}}
\def\righthook{{\vrule height5pt width0.4pt depth0pt}}
\def\leftrighthookfill{$\mathsurround=0pt \mathord\lefthook
     \hrulefill\mathord\righthook$}
\def\underhook#1{\vtop{\ialign{##\crcr$\hfil\displaystyle{#1}\hfil$\crcr
      \noalign{\kern-1pt\nointerlineskip\vskip2pt}
      \leftrighthookfill\crcr}}}
%
%


\def\hb{\hfill\break}
\def\ie{{\it i.e.\ }}
\def\eg{{\it e.g.\ }}

\def\ZZ{Z\!\!\!Z}               
\def\NN{I\!\!N}                 %
\def\RR{I\!\!R}                 
\def\CC{I\!\!\!\!C}             %

\def\bs{\bigskip}
\def\ss{\smallskip}



\def\state#1{|#1\rangle}


\nopagenumbers\pageno=0

\def\cC{{\cal C}} 
\def\cF{{\cal F}}
\def\lC#1{\cC^{(#1)}}

\def\fff{\cF(p^M,p^L)}

\def\whd{\widehat{d}}
\def\summ#1#2{\sum_{\scriptstyle #1 \atop \scriptstyle #2}}

\def\rel{{rel}} 
\def\bd{\overline{d}}

\def\whd{\widehat{d}}

\def\ara{\rightarrow}

\def\tphi{\tilde{\phi}}
\def\tI{\tilde{I}}

\baselineskip=0.8\baselineskip

\line{\hfill{CERN-TH.6346/91}}
\bigskip\bigskip
\centerline{{\bf GROUND RING FOR THE 2D NSR STRING}}
\bigskip\bigskip
\centerline{Peter Bouwknegt}
\smallskip
\centerline{{\it CERN - Theory Division}}
\centerline{{\it CH-1211 Geneva 23}}
\centerline{{\it Switzerland}}
\bigskip
\centerline{Jim McCarthy\footnote{$^{\dagger}$}{Supported by the NSF Grant
\#PHY-88-04561.}}
\smallskip
\centerline{{\it Department of Physics}}
\centerline{{\it Brandeis University}}
\centerline{{\it Waltham, MA 02254}}
\bigskip
\centerline{{ Krzysztof Pilch\footnote{$^\ddagger$}{Supported
in part by the Department of Energy Contract \#DE-FG03-84ER-40168.}}}
\smallskip
\centerline{{\it Department of Physics}}
\centerline{{\it University of Southern California}}
\centerline{{\it Los Angeles, CA 90089-0484}}
\bigskip
\bigskip
\centerline{{\bf Abstract}}\smallskip

We discuss the BSRT quantization of 2D $N=1$ supergravity coupled
to superconformal matter with $\hat{c} \leq 1$ in the
conformal gauge. The physical states are
computed as BRST cohomology.  In particular, we consider
the ring structure and associated symmetry algebra
for the 2D superstring ($\hat{c} = 1$).

\bs\bs
\centerline{Submitted to: {\it Nuclear Physics} {\bf B.}}
\vfil
\line{ CERN-TH.6346/91\hfill}
\line{ BRX TH-329 \hfill}
\line{ USC-91/38  \hfill}
\line{ December 1991\hfill}
\eject

\baselineskip=1.25\baselineskip
\footline={\hss\tenrm\folio\hss}


\def\NP{Nucl. Phys.\ }
\def\PL{Phys. Lett.\ }
\def\CMP{Commun. Math. Phys.\ }

\def\MPL{Mod. Phys. Lett.\ }

\def\LMP{Lett. Math. Phys.\ }

\bref\Pob{
A.M. Polyakov,  \PL {\bf 103B} (1981) 207.}
\bref\Po{
A.M. Polyakov,  \MPL {\bf A2} (1987) 893.}
\bref\Se{
N. Seiberg, Prog. Theor. Phys. Suppl. {\bf 102} (1990) 319.}
\bref\ED{
E. D'Hoker, \MPL {\bf A6} (1991) 745.}
\bref\IK{
I. Klebanov, Lectures at Spring School on String Theory and
Quantum Gravity, Trieste, 1991, PUPT-1271.}
\bref\LZb{
B.H. Lian and G.J. Zuckerman, \PL {\bf 254B} (1991) 417.}
\bref\LZe{
B.H. Lian and G.J. Zuckerman, \PL {\bf 266B} (1991) 21.}
\bref\BMPd{
P. Bouwknegt, J. McCarthy and K. Pilch, {\it BRST analysis of the
physical states for 2D gravity coupled to $c\leq 1$ matter},
CERN-TH.6162/91, to appear in \CMP}
\bref\Wi{
E. Witten, {\it Ground ring of two dimensional string theory},
IASSNS-HEP-91/51.}
\bref\KP{
I.R. Klebanov and A.M. Polyakov, {\it Interaction of discrete states
in two-dimensional string theory}, PUPT-1281.}
\bref\Da{
F. David,  \MPL {\bf A3} (1988) 1651.}
\bref\DK{
J. Distler and H. Kawai,  \NP {\bf B321} (1989) 509.}
\bref\DHK{
J. Distler, Hlousek and H. Kawai, Int. Jour. Mod. Phys. {\bf A5}
(1990) 391.}
\bref\BK{
M. Bershadsky and I. Klebanov, \NP {\bf B360} (1991) 559.}
\bref\IO{
K. Itoh and N. Ohta, {\it BRST cohomology and physical states in $2D$
supergravity coupled to $\hat{c}\leq1$ matter}, FERMILAB-PUB-91/228-T.}
\bref\BMPe{
P. Bouwknegt, J. McCarthy and K. Pilch, {\it BRST analysis of physical
states for $2D$ (super) gravity coupled to (super) conformal matter},
CERN-TH.6279/91.}
\bref\IMNU{
M. Ito, T. Morozumi, S. Nojiri and S Uehara,
Prog. Theor. Phys. {\bf 75} (1986) 934.}
\bref\Oht{
N. Ohta, Phys. Rev. {\bf D33} (1986) 1681;
{\it ibid.} Phys. Lett. {\bf 179B} (1986) 347.}
\bref\Shw{
J.H. Schwarz, Suppl. Prog. Theor. Phys. {\bf 86} (1986) 70.}
\bref\KS{
D. Kutasov and N. Seiberg, \PL {\bf 251B} (1990) 67.}
\bref\Hx{ M. Henneaux, \PL {\bf 183B} (1987) 59. }
\bref\Jose{
J.M. Figueroa-O'Farrill and T. Kimura,  \CMP {\bf 124} (1989) 105.}
\bref\LZc{
B.H. Lian and G.J. Zuckerman,  \CMP  {\bf 125} (1989) 301.}
\bref\FMS{
D. Friedan, E. Martinec and S. Shenker, \NP {\bf B271} (1986) 93.}
\bref\HMM{
G. Horowitz, R. Myers and S. Martin, \PL {\bf 218B} (1989) 309.}
\bref\BMPf{
P. Bouwknegt, J. McCarthy and K. Pilch, {\it Fock space resolutions
of the Virasoro highest weight modules with $c\leq1$},
CERN-TH.6196/91, to appear in \LMP {\bf 23}.}
\bref\Kutv{
D. Kutasov, {\it Some properties of (non) critical strings}, PUPT-1277.}
\bref\KMS{
D. Kutasov, E. Martinec and N. Seiberg, {\it Ground rings and their
modules in $2D$ gravity with $c\leq1$ matter}, PUPT-1293.}
\bref\BT{
R. Bott and L.W.\ Tu, {\it Differential forms  in algebraic
topology}, Springer-Verlag, New York, 1982.}
\bref\HS{
P.J. Hilton and U. Stammbach,
{\it A course in homological algebra}, Springer-Verlag,
New York, 1971.}
\bref\KLLSW{
V.A. Kosteleck\'y, O. Lechtenfeld, W. Lerche, S. Samuel and S. Watamura,
\NP {\bf B288} (1987) 173.}
\bref\Ka{
V.G. Kac, Lect. Notes Math. {\bf 94} (1979) 441.}


\newsec{ Introduction}

One success of the continuum approach to 2D gravity [\Pob,\Po]
(for a review, see [\Se--\IK] and references therein)
has been the computation of the physical states as BRST cohomology classes
[\LZb--\BMPd], at least for physical conditions under which the
worldsheet cosmological constant may be set to zero.
In particular, for the $c=1$ matter coupling, the BRST analysis constructs
in detail the infinite set of ``discrete states'' discovered by
matrix model calculations and continuum calculations of tachyon
scattering amplitudes.  The emergence of the interpretation of the
Liouville mode as the ``time'' coordinate of the embedding space --
and thus the identification of the model as two dimensional string
theory -- is particularly evident in the continuum approach.
More recently there has been some suggestion [\Wi,\KP] that
the model should be interpreted in terms of a three dimensional theory.

In [\Wi] this suggestion was based on the observation that
the discrete states at ghost number $-1$ provide a ring of
operators, in cohomology, which is characteristic of the model.
This ring can be identified as the polynomial ring generated by two
elements.  Further, the cohomology partners at ghost number $0$
give rise to spin $1$ currents which act as
derivations of this ring.  The algebra of
charges is just $\cW_{\infty}$ for one chiral sector.  When
left and right moving sectors are combined,
the total ring -- the ``ground ring'' --
is generated by four elements with
one constraint,
and thus defines a three dimensional space.
All this was discussed in [\Wi], together with a concrete identification
with matrix model results.

Soon after the conformal gauge quantization of 2D gravity models
was understood [\Da,\DK], the same ansatz was applied to 2D supergravity
coupled to superconformal matter with
$\hat{c} \leq 1$ ($\hat{c} = {2\over3}c$) [\DHK,\BK].
As models of string theory, the 2D NSR strings which arise for $\hat{c}=1$
are naturally of great importance.  There is apparently no
matrix model formulation available, which makes application
of the continuum approach extremely relevant in this case.
The BRST analysis of the physical spectrum
was recently carried out [\IO], and also
discussed in [\BMPe] (which contains the
further projection onto $\hat{c}<1$ minimal models as well).

For a given $N=1$ super Virasoro module ${\cal V}$, the
constraints $T(z) \sim 0 \,, G(z) \sim 0$ (from coupling to supergravity)
can be implemented by the BRST operator
\eqn\PBg{
d = \oint {dz\over 2\pi i} :\! \left( c(z)(T(z) +\half T^G(z)) +
 \ga(z)(G(z) + \half G^G(z)) \right)\! : \,,
}
acting on the tensor product module ${\cal V}\otimes \cF^G$.
The operators $T^G(z)$ and $G^G(z)$ generate the $N=1$ superconformal
algebra on $\cF^G$, which is the tensor product of the Fock space of the
spin $(2,-1)$ $bc$-ghosts with that of the spin $({3\over2},-\half)$
$\be\ga$-ghosts.  The BRST operator $d$ is nilpotent
provided the central charge of ${\cal V}$ is
equal to $10$ [\IMNU--\Shw].
For the 2D NSR string the super Liouville system
consists of a free fermion together with a free scalar with
background charge, and $\cal V$ is the product
of the corresponding Fock spaces with those of the free matter system at
$\hat{c}=1$.  More generally, the computation can be done for the
case that both systems have a background charge.  The minimal model
matter coupling is then obtained by projection.

In this paper we continue this program by discussing the
structure of the cohomology in the 2D NSR strings.
That is, we determine the ring structure, and the corresponding
symmetry algebra of currents, for the ``critical'' case in which
the cosmological constant vanishes.  We follow the approach in [\Wi],
and in fact the chiral structure turns out to be almost precisely the
same as the bosonic case discussed there.  Although, as we show,
this could be anticipated by a ``kinematic analysis,'' a detailed
calculation is actually required to establish the result.
The ground ring structure depends on how the left and right
moving sectors are joined.  For the
NSR string we can simply proceed with the full theory, and
thus produce a 2D NSR closed string with the same ground ring structure
and symmetry as the 2D bosonic string.  Alternatively we may
enforce a consistent GSO-type projection of the model
[\KS], which gives a restriction of the
ground ring and its symmetries.

These results are presented as follows.  In Section 2 we summarize and
further discuss the BRST cohomology computed for
2D world sheet $N=1$ supergravity coupled to a free supermultiplet
with background charge.  This contains directly the results required
for the 2D NSR string, which are then applied in Section 3
to obtain the ring structure and the algebra of charges for
one chiral sector -- both for the 2D NSR string, and its
GSO projection.  Putting together both left and right
movers in Section 4, we derive the ground rings of these models and
discuss their corresponding symmetry algebras.  We have gathered
all conventions and notations used herein into Appendix A.
Further, we review in Appendix B the details of the computation
of the relevant BRST cohomology, both the relative and absolute cohomologies
for both NS and R sectors.  For convenience, Appendix C contains various
details about the bosonization required for writing the
explicit representatives used in the text.  Finally, in Appendix D,
we present the generalization to the superconformal case
of results on the structure of the $c=1$ (now $\hat{c}=1$)
Fock space -- in particular we introduce ``super-Schur polynomials,''
in terms of which expressions for the
singular vectors may be given.


\newsec{ Summary of the  BRST cohomology calculation}

\def\Ker{{\rm Ker}\,} 

In this section we summarize the cohomology of the BRST operator $d$
given in \PBg, where the module ${\cal V}=\fff$ is taken as the tensor
product of
Fock spaces $\cF(p^M,Q^M)$ and $\cF(p^L,Q^L)$ corresponding to two
scalar supermulitplets $(\phi^M,\psi^M)$ and $(\phi^L,\psi^L)$ with
background charges $Q^M$ and $Q^L$, respectively.  The condition of
nilpotency then implies that \footnote{$^1$}{Note that in our
conventions both $p^L$ and $Q^L$ are purely imaginary.} \eqn\JIb{
(Q^M)^2 + (Q^L)^2 =-1 \,.} The resulting cohomology spaces will be
denoted $H_{abs}^{(n)}(\fff,d)$.  A summary of notations and
conventions is given in Appendix A, and for completeness we include
technical details of the computations in Appendix B.

The BRST charge $d$ decomposes under ghost zero modes in the two
sectors as \eqnn\JKa\eqnn\JKb
$$\eqalignno{
{\rm NS:} \quad\quad d&= L_0c_0
- Mb_0 + \whd\,, & \JKa \cr
{\rm R:} \quad\quad d&= L_0c_0 -
(M+\ga_0\ga_0)b_0 + \bd & \cr
  &= L_0c_0 - (M+\ga_0\ga_0)b_0 + (F\ga_0 + N\be_0  + \whd\,) \,. &
\JKb\cr}
$$
where none of the operators (besides $\bd$) in the expansion contains ghost
zero modes.  We have $\{d,b_0\} = L_0 = L_0^M +L_0^L + L_0^G$ and
$[d,\be_0]=F-2\ga_0b_0 = G_0 = G_0^M + G_0^L + G_0^G$.  The oscillator
expressions for these operators can be read off from the formulae
given in Appendix A, and the nontrivial
(anti-) commutators between them, following from $d^2=0$,  are
\eqnn\commutata\eqnn\commutatb
$$\eqalignno{
{\rm NS:} \quad\quad \whd{\,}^2 & = ML_0 \,, & \commutata \cr
{\rm R:} \quad\quad FF & =L_0\,,\quad [F,M]=2N\,,\quad \whd{\,}^2=
ML_0 + NF\,. & \commutatb \cr}
$$

The computation of the BRST cohomology proceeds in two steps.  First
one determines the relative cohomology, denoted
$H_{rel}^{(n)}(\fff,d)$, which is the cohomology of $d$ on the
subspace for which $d$ reduces to $\whd$.  For NS this subspace is (as
for the bosonic case)
\eqn\KPspac{
\cF_\rel(p^M,p^L)\equiv\fff \cap \Ker\,
L_0 \cap \Ker\, b_0\,,}
whilst for R we must restrict to
\eqn\KPsccs{
\cF_\rel(p^M,p^L)\equiv\fff \cap \Ker\,
G_0 \cap \Ker\, b_0 \cap \Ker\, \be_0 \,.}
In the R sector we will also distinguish
additional subspaces of $\fff$, namely \eqnn\KPexspc\eqnn\KPexsps
$$\eqalignno{{\cal K}(p^M,p^L)&\equiv\fff\cap\Ker L_0\cap\Ker b_0\,,&
\KPexspc\cr \cF^{L_0}(p^M,p^L)&\equiv {\cal K}(p^M,p^L)\cap
\Ker\be_0\,.&\KPexsps\cr}$$ Note that $\bd$ (defined in \JKb\ )
is nilpotent on ${\cal K}(p^M,p^L)$.

The resulting relative cohomology is summarized as follows:
Parametrize $(p^M,p^L)$ by $r,s\in\CC$
\eqn\JOa{
\eqalign{
p^M-Q^M & = \sqrt\half
(r\al_+ + s\al_-)\,, \cr i(p^L-Q^L) & = \sqrt\half (r\al_+ - s\al_-)
\,,\cr}
}
where
\eqn\PBm{
\al_\pm = \sqrt\half (Q^M\pm i Q^L) \,.}
Then $ H^{(n)}_{rel}(\fff,d)$ is nontrivial only
in the following four cases [\IO,\BMPe] \smallskip

\item{(i)} If both $r=s=0$ (\ie $p^M=Q^M$ and $ip^L=iQ^L$), then
$$H^{(n)}_{rel}(\fff,d) = \cases{ \CC \oplus \CC & if $n=0$
and $\ka = 0\,,$ \cr \CC & if $n=0$ and $\ka = \half \,,$ \cr
0 & otherwise. \cr} $$
\item{(ii)} If either $r=0$ or $s=0$ (\ie $\De(p^M)+\De(p^L) = \half$),
then $$H^{(n)}_{rel}(\fff,d) = \cases{ \CC & if $n=0\,,$ \cr 0 &
otherwise. \cr} $$
\item{(iii)} If $r,s\in\ZZ_+\,, r-s\in2\ZZ+(1-2\ka)$,
then $$ \eqalign H^{(n)}_{rel}(\fff,d) = \cases{ \CC & if $n=0,1\,,$ \cr
0 & otherwise .\cr} $$
\item{(iv)} If $r,s\in\ZZ_-\,,
r-s\in2\ZZ+(1-2\ka)$, then $$ H^{(n)}_{rel}(\fff,d) = \cases{ \CC & if
$n=0,-1\,,$ \cr 0 & otherwise . \cr} $$ \smallskip

Note that the ``discrete states" arise only if both  $\cF(p^M,Q^M)$
and $\cF(p^L,Q^L)$ are reducible.  They occur at the same level,
namely ${rs\over2}$, as the null vectors in these modules.

Given this result for $H_{rel}^{(n)}(\fff,d)$, the full cohomology may
be determined.  Indeed, for the NS case, arguments as in
[\BMPd] show that the cohomology is simply doubled due to the ghost
zero mode $c_0$. More precisely, each  relative
cohomology state $\psi$ gives rise to two absolute cohomology states,
\eqn\JRe{
\psi^{(1)}=\psi \quad\quad \psi^{(2)}=c_0\psi+\phi \,,
}
where $\phi$ is a solution (which always exists) to the equation
$M\psi=\whd\phi$.

In the R sector, as in the ten dimensional NSR string
[\Hx--\LZc], the situation
is more complicated but the final result is the same -- we just have a
doubling of states.  For the exceptional case, when $p^+=p^-=0$,
the most convenient approach is simply to
enumerate all the states annihilated by $L_0$ and  find
the explicit form of the BRST operator acting on them.
In the NS sector we verify immediately that the two states, which are also
the absolute cohomology representatives, are $\state{Q^M,iQ^L;-1}$ and
$c_0\state{Q^M,iQ^L;-1}$.  In the R sector,
such states are linear combinations of the following basis states
\eqn\gogogo{
(\ga_0)^{m_+}\state+ \,,\quad (\ga_0)^{m_-}\state -\,,
\quad (\ga_0)^{n_+}c_0\state+ \,,\quad (\ga_0)^{n_-}c_0\state -\,,}
where $m_\pm\,,n_\pm\geq0$, and $\state-$ and $\state+$ denote
the two degenerate vacua $\state{Q^M,iQ^L;-\half}$
and $\psi_0^+ \state{Q^M,iQ^L;-\half}$, respectively.
Since for $p^+=p^-=0$ the fermion zero mode terms are absent in $G_0$,
these two vacua span the relative cohomology. The full BRST operator
acting on the states in \gogogo\ is simply $d=-\ga_0\ga_0 b_0$,
so one easily sees that the absolute cohomology is doubly degenerate at
ghost number 0 and 1, and is generated by $(\ga_0)^{m_\pm}\state\pm$
with $m_\pm=0,1$.

For all other cases,
the reason that there is no infinite degeneracy due to the bosonic zero
mode $\ga_0$ goes back to the simple fact that
in $\cF^{L_0}(p^M,p^L)$, any state in ${\rm Ker}\, G_0$ is in fact in
${\rm Im}\, G_0$. [Note
that on this space  $G_0$ and $F$ coincide.]  The complete
calculation of this result is deferred to Appendix B, here
we simply outline the procedure.
The two absolute cohomology states corresponding to a given relative
cohomology state  $\psi_0$ are:  $\psi^{(1)}=\psi_0$  and
$\psi^{(2)}=\phi_0 + \ga_0\phi_1+c_0\psi_0$, where
$\phi_0,\phi_1\in\cF^{L_0}(p^M,p^L)$ are solutions to the following
system of equations
\eqn\stuff{\eqalign{
G_0\phi_1-\psi_0=&0\,,\cr
G_0\ph_0+\whd\ph_1=&0\,,\cr \whd\ph_0-N\ph_1-M\psi_0=&0\,,\cr}
}
equivalent to $d\psi^{(2)}=0$. Let us verify that this system
 has a solution for  any relative cohomology state  $\psi_0$.  Since
$G_0\psi_0=0$, we can always solve the first one and determine $\phi_1
$. Then  $G_0\whd\ph_1=0$, and  we may determine $\phi_0$ using the
second equation. In fact both $\phi_0$ and $\phi_1$ are determined only
up to $G_0 $  exact terms.
Using this freedom, we may rewrite the third equation
as the statement that there exists $\rh \in \cF_\rel(p^M,p^L)$ such that
\eqn\rewrit{\whd\ph_0-N\ph_1-M\psi_0= \whd\rh \,.}
Now, using the first two equations in
\stuff\ and identities in \commutatb\ we verify that the l.h.s.\ is
annihilated both by $G_0$ and $\whd$, and thus is an element of
relative cohomology.  However, since there can be no
relative cohomology at ghost number (gh)$(\psi_0)+2$, we conclude that
a $\rh$ satisfying \rewrit\ must exist.

Of course, a more detailed
analysis is still required to show that the states $\psi^{(1)}$ and
$\psi^{(2)}$ are indeed nontrivial and that linear combinations of such
states exhaust all of the absolute cohomology.  This is discussed in
detail in Appendix B.

Some comments are in order.

\noindent 1. The above results give the
cohomology in a particular picture for the
$\be\ga$ system; namely $Q_{3/2}= -\half - \ka$.  However, as was realized
in [\FMS] and clearly explained in [\HMM], there exists isomorphic
copies in any picture differing by integer charge.  The isomorphism is
given explicitly by repeated action of the zero mode $X_0$ of the picture
changing operator, and the zero mode $Y_0$
which is the inverse to $X_0$ in cohomology.  This
is {\it not} generally an isomorphism in relative cohomology; although
it is true that $[b_0,X_0]=0$, this is not so for $Y_0$.  The
picture changing isomorphism will be essential in Section 3.

\noindent 2. The projection to the $\hat{c} < 1$
superconformal minimal models is obtained via an appropriate
free field resolution of the irreducible representations, as
detailed in [\LZb,\BMPd,\BMPf,\BMPe].

\noindent 3. It is well known that the ${\rm gh} =0$ representatives
for the 2D bosonic string can be written in the form of ``$c=1$
singular vector $\times$ Liouville vacuum.''  The singular
vectors, in turn, have known expressions in terms of Schur polynomials.
The analogous result holds for the 2D NSR string, but now
in terms of ``super" Schur polynomials.
We detail the construction in Appendix D.


\newsec{ The chiral structure of the $\hat{c} = 1$ model}

For the remainder of the paper we will restrict our attention to the
case $\hat{c} = 1$, where $iQ^L=1$ and thus
$\al_\pm = \pm 1/\sqrt2$,
which we may think of as the 2D NSR string.
We will exploit the analysis of Section 2, together with
standard results of conformal field theory, to obtain some
insight into the structure of this model.

It was recently
established for the 2D bosonic string that the operator
cohomology contains a natural ring structure, and that
relative cohomology states at ${\rm gh}=0$ give rise to a symmetry algebra
which acts as derivations of the ring [\Wi].  Specifically, to a given
$\psi \in H^{(-1)}_{abs}(\fff,d) \,\, (\simeq H^{(-1)}_{rel}(\fff,d))$
we may associate an $L_0$-weight zero operator $\Psi(z)$ such that
\eqn\JTa{
\psi = \lim_{z \to 0}\Psi(z) |0\rangle \,,
}
where $|0\rangle$ denotes the $sl(2,\RR)$ vacuum of the model.
Given two such, $\Psi_1(z)$ and $\Psi_2(z)$, the product
\eqn\JTb{
(\Psi_1\Psi_2)(z) \equiv {1\over 2\pi i}
\oint_z {dx\over{x-z}} \Psi_1(x)\Psi_2(z)
}
is commutative and associative for the states of the bosonic string --
indeed all singular terms in the OPE are $d$-trivial by standard arguments
[\Wi].
Further, from a representative $\psi \in H^{(0)}_{rel}(\fff,d)$ we may
derive a spin one current; it is just defined as that operator $J(z)$
associated to $b_{-1}\psi$.
We write
\eqn\JTg{
J(z) = (b_{-1}\Psi)(z) \,.
}
The corresponding charge is BRST invariant.  Amongst other things, it
was shown in [\Wi] that the ring can be identified with the polynomial ring
generated by two elements, $x$ and $y$ say,  on which the charges
are represented as vector fields.  The problem was greatly simplified by the
fact that two of the symmetry charges act as $\p_x$ and $\p_y$.

This product, and the definition of symmetry current,
may be immediately taken over to the NSR string, and we will
soon consider just how much of the other structure does also.  First, however,
we may expect to gain some understanding by a simple examination
of the ``kinematics'' involved.  From the results of
Section 2 we see that the
representatives of $H^{(-1)}_{abs}(\fff,d)$ are parametrized by
two negative integers.  The element $\psi_{(r,s)}$, at level ${rs\over2}$,
has momenta
\eqn\JTc{
[p^M,ip^L] = [\half(r-s), \half(r+s+2)] \,.
}
Half the states, those with $r-s \in 2\ZZ$, appear in the NS sector, while
those with $r-s \in 2\ZZ + 1$ come from the R sector.
By adding the momenta, we clearly find
\eqn\JTd{
(\Psi_{(r,s)}\Psi_{(r',s')})(z) \sim \Psi_{(r+r'+1,s+s'+1)}(z) \,,
}
where $\sim$ emphasizes that the identification on the r.h.s. is forced, but
only once it is established that the r.h.s. is nonzero.  It is now clear that,
with this proviso, the entire chiral ring is generated by the operators
$x \equiv \Psi_{(-1,-2)}$ and $y \equiv \Psi_{(-2,-1)}$,
with momenta $[\half, -\half]$ and $[ - \half, -\half]$ respectively,
which arise in the R sector.  For
\eqn\JTe{
\Ps_{(r,s)} \sim x^{-s-1}y^{-r-1}\,,\qquad r,s<0\,,
}
and thus ``kinematically'' one obtains all possibilities.

Representatives of $H^{(0)}_{rel}(\fff,d)$ are parametrized by
two numbers $r,s\in\CC$, with either $r=0$, $s=0$ or
both $r,s\in\ZZ$ with $rs>0$ (see section 2). We will restrict our attention to
the subset parametrized by integers $r,s \leq 0$, since the remainder of
the associated currents \JTg\ will act trivially on the ring, as
follows (a posteriori) by kinematical reasoning as in [\Wi].
{}From these states $\psi_{(r,s)}$, we construct the spin $1$
current $J_{(r,s)}$ as in \JTg.  The kinematics of these operators
is exactly as in equation \JTd\ above,
so we may, for example, immediately deduce
that the only candidate for $\p_{x}$ is $\p_{x} \sim J_{(-1,0)}$.
For, it is the only current which acts on $x$ to produce
a state with the momenta $[0,0]$ of the identity operator.
Similarly we have the candidate $\p_{y} \sim J_{(0,-1)}$.
So far, the combined NS and R structure does not look much different from
the one that occurred in the bosonic case [\Wi].

To make more concrete statements requires some calculation, but first
we should clear up several points which may be troubling the reader at this
stage.  The cohomology calculation in Section 2 was done in a particular
picture, and thus a more precise notation for the ring elements
and currents above
is $\Psi_{(r,s)}^{(q)}$ and $J_{(r,s)}^{(q)}$, where $q = -\half - \ka$,
$\ka = 0, \half$ depending on which sector
the operators arise from.  However, now
the result of the product of operators from pictures $-\half - \ka_1$ and
$-\half - \ka_2$ will be an operator in the $(-1 - \ka_1 - \ka_2)$ picture.
For $\ka_1 = \ka_2 = 0$ we straightforwardly have the product
R$\times$R$\ara$NS, but otherwise we go outside the set $q=-\half , -1$.
Fortunately this is not a problem, since -- as emphasized in Section 2 --
picture changing provides an isomorphism in cohomology.
By standard contour deformation arguments one can verify that
picture changing is, in fact, a ring isomorphism.
Thus we
may simply identify all states which are related by picture
changing, and the operator cohomology still defines a ring
modulo this identification.
The remaining products are then
NS$\times$NS$\ara$NS, NS$\times$R$\ara$R, and R$\times$NS$\ara$R.
In the above notation we easily find for example

\eqn\RIa{
\eqalign{
\psi_{(-1,-2)}^{(-{1\over2})} & =
\left( b_{-1} - \sqrt\half \be_{-1}\ps_0^+ \right)
\state{\half,-\half;-\half} \,, \cr
\psi_{(-2,-1)}^{(-{1\over2})} & =
\left( \be_{-1} + \sqrt2 b_{-1}\ps_0^+ \right)
\state{-\half,-\half;-\half} \,, \cr
\psi_{(-1,-3)}^{(-1)} & = \left( \al_{-1}^-\be_{-{1\over2}} +
b_{-1}\ps^-_{-{1\over2}}
-\sqrt2 \be_{-{3\over2}} \right) \state{1,-1;-1} \,, \cr
\psi_{(-3,-1)}^{(-1)} & = \left( \al_{-1}^+\be_{-{1\over2}} +
b_{-1}\ps^+_{-{1\over2}}
+\sqrt2 \be_{-{3\over2}} \right) \state{-1,-1;-1} \,, \cr
}
}
and upon bosonizing we have
\eqn\JJa{
\eqalign{
x & =  [S^-e^{-{i\over2}\tphi} - {1\over{\sqrt{2}}}S^+ \p \xi
e^{-{3\over2}i\tphi} c]e^{{i\over2}(\phi^M + i\phi^L)} \,, \cr
y & = [S^- \p \xi e^{-{3\over2}i\tphi} c -
\sqrt{2} S^+ e^{-{i\over2}\tphi}] e^{-{i\over2}(\phi^M - i\phi^L)} \,,  \cr
\Ps_{(-1,-3)}^{(-1)} &
= \left( -c\sqrt{2}\p^2\xi e^{-2i\tphi} - ci\p\tphi
\p\xi e^{-2i\tphi} + \psi^-e^{-i\tphi} + i\p\phi^-
\p\xi e^{-2i\tphi} c \right) e^{i(\phi^M+i\phi^L)} \,, \cr
\Ps_{(-3,-1)}^{(-1)} &
= \left( c\sqrt{2}\p^2\xi e^{-2i\tphi} + ci\p\tphi
\p\xi e^{-2i\tphi} + \psi^+e^{-i\tphi} + i\p\phi^+
\p\xi e^{-2i\tphi} c \right) e^{-i(\phi^M-i\phi^L)} \,. \cr}
}
This still leaves a question for the currents, however, since these have
been defined via the action of $b_{-1}$ in a given picture.  We may
define the currents in the $q$ picture by
\eqn\JTh{
J_{(r,s)}^{(q)}(\Psi^{(q')}) = J_{(r,s)}^{(q-1)}(\Psi^{(q'+1)}) \,,
}
where $\Psi_{(r,s)}^{(q)}$ corresponds to a state in $H^{(0)}(\fff,d)$
picture changed to the $q$  picture as discussed above.
To see that this is sensible, it is sufficient to note the identity
\eqn\JTi{
\oint_0 dz (b_{-1}(X_0\Psi_{(u,v)}^{(q-1)}))(z)(\Psi^{(q')})(0) =
\oint_0 dz (b_{-1}\Psi_{(r,s)}^{(q-1)})(z)(X_0\Psi^{(q')})(0) \,.
}
So, in fact, we have
\eqn\JTz{
J_{(r,s)}^{(q)}(z) = (b_{-1}\Psi_{(r,s)}^{(q)})(z) \,.
}
A last comment along these lines is that, since it is always possible to
choose a representative $\psi^{(-1)}$ of $H^{(0)}_{rel}(\fff,d)$ which is a
superconformal highest weight (and ghost vacuum) as we show
in Appendix D, one might try to introduce spin 1 ``superpartners" to the
currents as, for example,  $\widetilde{J} \equiv \be_{-1/2}G_{-1/2}\psi$.
This obeys all the required properties in the same way as the usual current.
However, explicit computation shows that there are no new currents here.

With this in hand, we may construct some examples of the currents
in this notation,
\eqn\JTk{
\eqalign{
J^{(-{1\over2})}_{(-1,0)} & = S^- e^{-{i\over2} \tphi}
e^{-{i\over2}(\phi^M + i\phi^L)}\,, \cr
J^{(-{1\over2})}_{(0,-1)} & = S^+ e^{-{i\over2} \tphi}
e^{{i\over2}(\phi^M - i\phi^L)}\,, \cr
}
}
\eqn\JTj{
\eqalign{
J_+(z) & \equiv J^{(0)}_{(0,-2)}(z) = \left( \psi^M \right)e^{i\phi^M} \,. \cr
J_0(z) & \equiv J^{(0)}_{(-1,-1)}(z) = \left( i \partial\phi^M \right)\,, \cr
J_-(z) & \equiv J^{(0)}_{(-2,0)}(z) = \left( \psi^M \right)e^{-i\phi^M}\,, \cr
}
}

Let us now consider how well the kinematical predictions are borne out
by explicit computation.  Most important is the identification
of $\p_{x}$ and $\p_y$, which we can check from \JTk\ and \JJa\ , \eg
\eqn\JJc{
J^{(-{1\over2})}_{(-1,0)}(z)x(w) \sim {1\over{(z-w)}}{1\over{\sqrt{2}}}
\p_w \xi e^{-2i\tphi} c(w) \,.
}
The residue of the pole is easily identified as the
picture changed representative of the identity in the $q=-1$ picture.
Similarly we may check that $\p_{y}$ is correctly identified.
By this calculation we have verified (in exactly the same way as
the analogous statement was proven in [\Wi]) that indeed the
cohomology ring is precisely the polynomial ring generated
by the two elements $x$ and $y$!  Thus we immediately
determine that for $r,s >0$, $J_{(r,s)}$ act trivially on the
ring -- since they map the generators to states which vanish
in cohomology.

In fact, as perhaps one
could also anticipate from the kinematic analysis, the chiral
structure for the NSR string is almost exactly parallel to
that of the bosonic string.  In particular, the currents
$J_+,J_-$ and $J_0$
defined in \JTj, with $r+s=-2$, generate an $sl(2)$ algebra
under which $x$ and $y$ transform as a doublet.
Indeed the complete spectrum may be decomposed under it (see also the
discussion in Appendix D).
States with the same Liouville momentum are in the
same multiplet (from the kinematics \JTd, $(r'+s') \ara (r'+s')$ when acted
on by a current with $r+s = -2$).
Thus the symmetry currents occur in $sl(2)$ multiplets with spin $j=|r+s|/2$,
the highest weight in the multiplet always being one of
the ``discrete tachyon" vertices, $J_{(0.-2j)}$.
The ring decomposes into $sl(2)$ multiplets with spin $j=(|r+s|-2)/2$.
One may now check that the arguments in [\Wi] go through
for the nontriviality of the action of $J_{(r,s)} \,\,, r,s \leq 0$,
on the ring, and the identification of the symmetry algebra with
$\cW_{\infty}$.  Indeed this action is just proportional to that
of the area preserving polynomial vector fields on the $x-y$ plane;
\ie, given the ``hamiltonian'' $h(x,y) = x^{-s}y^{-r} \,,
r,s \leq0$, $J_{(r,s)}$ acts on the
ring as the vector field
\eqn\APa{
\p_yh(x,y){\p\over{\p x}}
- \p_xh(x,y){\p\over{\p y}} \,.
}

There are, however, two last remarks which should be made regarding this
analysis of one chiral sector. The first is that in the above analysis we
have made a particular choice of cocycles for bosonizing the fields
such that the ring product is commutative.  In general it
is ``phase commutative,'' in the sense that the opposite orders of product
agree when a well-defined relative phase is included.
We give a brief outline of this choice in Appendix C.
One easily shows that associativity is still
guaranteed, so we do not believe there is any physical distinction
between different choices.  In particular,
one may show by direct calculation that $x^2 \sim \Ps_{(-1,-3)}$ and
$y^2 \sim \Ps_{(-3,-1)}$, completely independent of the choice of cocycles.

The second remark is that there is a well-defined ``GSO projection''
from which an analogue of the superstring may be defined [\KS].
Motivated by the corresponding analysis of compactified critical NSR strings,
this is imposed by restricting to the set of operators which are local
with respect to the spacetime supersymmetry charge.  For the
2D NSR string, because of the background charge required in the
Liouville sector, spacetime translation invariance is ruined along with
the usual supersymmetry algebra, and the analogue operator is the
spin 1 current
\eqn\SSa{
Q_S(z) = e^{-{i\over2}\tphi}S^-e^{i\phi^M}\,,
}
which defines a {\it nilpotent} charge.    [Note that this is obtained
by acting with $b_{-1}$ as before, but on the cohomology state in the
Ramond sector with $r=0 \,, s=-2$.]
However, it may still be used to make the
projection [\Kutv], and the resulting projected ring is just the restriction
to even powers of $y$, with no restriction on the power of $x$.  The
induced symmetry currents are those preserving this restriction.  From
\APa\ we are clearly left with those currents corresponding to hamiltonians
which are odd powers of $y$ -- with again no restriction on the $x$ power.
All this may be summarized by introducing the variable $z=y^2$.  The
GSO projected chiral ring is just the polynomial ring generated by $x$
and $z$.  The symmetry algebra of this ring is the subset of the
area preserving vector fields on the $x-y$ plane which map the
GSO projected ring to itself.  This constraint implies the hamiltonians
of the subalgebra are just those of the form
$x^{-s}y^{-2r+1} \,, r,s \leq0$, and the corresponding
vector fields on the $x-z$ plane are obtained as in \APa\ .
One sees immediately that these are {\it not} area preserving maps
of the $x-z$ plane -- or, more precisely, they preserve the area
in the metric induced by the transformation $x \ara x \,, y \ara z=y^2$.
It should also be noted that $Q_S$ acts trivially on the
projected ring, simply by kinematic reasoning, so there are
no spacetime fermionic elements in this ring.


\newsec{Conclusions: the ground rings and symmetries}

For the 2D NSR closed string we must now join left and right
sectors to obtain the physical theory.  From the
discussion of Section 3 we see that the analysis follows precisely
that of the bosonic string given in [\Wi].  For completeness we sketch
those results in the present context.  From
the NSR string one can make another interesting string theory -- the
2D superstring -- by imposing the GSO projection.  We will also
discuss the structure of this theory.

The ground ring is obtained by putting together the left moving
and right moving (distinguished by a $'$) rings with the
constraint that $p^L = p^{L}{}'$.  For the 2D NSR string one
immediately finds that
the ground ring  -- at the ``critical" point where the cosmological constant
vanishes -- is just the ring of polynomial functions on the quadric
cone $Q$ defined by
\eqn\Coa{
a_1a_2-a_3a_4 = 0\,,
}
where the generators are
\eqn\Cob{
\eqalign{
a_1 & = xx'\,, \cr
a_2 & = yy'\,, \cr
a_3 & = xy'\,, \cr
a_4 & = yx' \,. \cr
}
}
Note that we have allowed the matter scalar $\phi^M$ to be
compactified.  For the uncompactified string we have the further
constraint that $p^M=p^M{}'$, and the ground ring reduces to that
generated by just $a_1$ and $a_2$.

The quantum symmetry currents are the spin (1,0) and (0,1) operators
with $sl(2)$ spin $j \in \half\NN$
\eqn\Joc{
\eqalign{
J_{j,m,m'} &= J_{(-j-m,-j+m)}\Psi'_{(-j-m',-j+m')}\,, \cr
J'_{j,m,m'} &= \Psi_{(-j-m,-j+m)}J'_{(-j-m',-j+m')}\,, \cr
}
}
where $-j \leq m \leq j \,, -j+1 \leq m' \leq j-1$.
Again note that the constraint $p^L=p^L{}'$ has been imposed.  The
resulting symmetry algebra is just the diffeomorphisms generated by the
volume preserving polynomial vector fields on the quadric $Q$.
When the matter sector is uncompactified, we must further restrict to those
currents which preserve the matching $p^M=p^M{}'$.  One is left with
those area preserving vector fields on the $a_1 - a_2$ plane which
preserve  the locus $a_1a_2=0$.

All this has simply been a copy of the bosonic string [\Wi].
We may now project to the 2D superstring by restricting
to those operators which are local with respect to the
``spacetime supersymmetry"
current, applied to each chiral sector.  As we saw in Section 3,
the chiral ring is then generated by $x$ and $z=y^2$, and on putting
them together we find a polynomial ground ring generated by
\eqn\Jod{
\eqalign{
b_1 & = xx'\,, \cr
b_2 & = zz'\,, \cr
b_3 & = x^2z' \,,\cr
b_4 & = zx'^2 \,, \cr
}
}
which satisfy the constraint
\eqn\Joe{
b_1^2b_2-b_3b_4 = 0 \,.
}
One may check that the quantum symmetry algebra, obtained analogously
to \Joc\ from the chiral algebra of the GSO projected chiral rings,
acts on this new subspace as vector fields.

There are some simple observations which can be made in conclusion.
The ``energy operator"
$i\p\phi^M$ is exactly the current $J_0$ of \JTj, so it appears that
the arguments in [\Wi] do not characterize any difference between the
matrix model description of the 2D bosonic string and an analogous
description in the NSR case -- if it exists.
In fact the real distinction in BRST structure
seems to only be in the doubling of relative cohomology  when $r$ or
$s$ vanishes (in the notation of Section 2), where the constraint on
the difference $r-s$ is lifted.  It is worth noting that the
energy operator clearly survives the GSO projection.
This is true even if the
GSO projection is defined by one of the other Ramond operators
with one of $r$ or $s$ vanishing, and $r-s \in 2\ZZ$.  [The
choice used in the present paper is most natural within the
class of extended models discussed in [\KS,\Kutv], where our
analysis could also be applied.]  Finally, for the GSO projected ring, the
symmetry charges act as vector fields which must preserve the volume of the
metric induced from that on the quadric,
$\Theta = db_1{{db_2}\over{\sqrt{b_2}}}{{db_3}\over{b_3}}$.
The significance of these remarks for understanding 2D superstrings
is yet to be determined.
\bs


\noindent
{\it Note added:} While preparing this manuscript we received a paper by
Kutasov, Martinec and Seiberg [\KMS], in which they discuss consequences of
the existence of ring structure in 2D string models.  They also argue that
the ground ring of the uncompactified 2D NSR string is as we determined
above.


\appendix {A} {Notations and conventions}

By scalar supermultiplet with background charge we mean the
pair $\phi(z)$, a free scalar field with background charge, and
$\psi(z)$, a free spin $\half$ real fermion, with
fundamental two-point functions
\eqn\LZac{
\vev{ \ph(z)\ph(w) }  = - \ln (z-w)\,,\qquad \vev{ \psi(z)\psi(w) } =
{1\over{(z-w)}} \,.
}
We take the following conventions for the $N=1$ superconformal currents
\eqn\PBc{
\eqalign{
T(z) & = -\half :\p\ph\p\ph: + iQ\p^2\ph - \half \ps\p\ps\,, \cr
G(z) & = i\p\ph\ps + 2Q\p\ps \,,\cr
}}
so that the central charge is given by
\eqn\JIa{
\hat{c} = \textstyle{{2\over3}}c = 1-8Q^2 \,.
}
The half-integer-spin fields can be consistently subjected
to different boundary conditions, leading to two distinct sectors which
we parametrize by $\ka = \half$ (Neveu-Schwarz) and $\ka = 0$ (Ramond).
In terms of modes,
$i\p\ph(z) = \sum_{n\in\ZZ} \al_nz^{-n-1}$, ${p=\al_0}$,
and $\psi(z) = \sum_{m\in\ZZ+\ka} \psi_mz^{-m-1/2}$,
and we have
\eqn\LZd{
[\al_m,\al_n] = m\de_{m+n,0}\,,\qquad \{\psi_m,\psi_n\} = \de_{m+n,0}\,.
}
Throughout the paper operators are always normal ordered with respect to
the $SL(2,\RR)$ vacuum (which is contained in the NS sector).
We will denote the Fock space built on the vacuum state $|p\rangle$ with
momentum $p$ by $\cF(p)$.  The conformal dimension of the corresponding
Virasoro representation is
\eqn\KPvb{
\De(p)=\half p(p-2Q) + \textstyle{{1\over16}}(1-2\ka) \,.
}
In the text we distinguish between the Liouville and matter
fields by writing superscripts $L$ and $M$ respectively.  The
total fermion number charge will be denoted $Q_{1/2}$.

The $N=1$ superconformal algebra of the combined ghost system is generated
by
\eqn\PBe{
\eqalign{
T^G(z) & =\ :\! c(z)\p b(z) + 2\p c(z) b(z) - \half \ga(z)\p\be(z) -
\textstyle{{3\over2}} \p\ga(z)\be(z)\! :\,, \cr
G^G(z) & = -2b(z)\ga(z) + c(z)\p\be(z) + \textstyle{{3\over2}}
\p c(z)\be(z) \,. \cr
}}
The choice of NS or R sector for the bosonic ghosts must be the same as
that for $\psi^M$ and $\psi^L$.

The ghost number current of the spin $\la$ pair (``fermion number'' if
$\la = \half$) is denoted $j_{\la}$,
and its charge by $Q_{\la}$.
We will use the notation $|p^M,ip^L;q\rangle$ for vacuum states,
where $q$ denotes the ``picture'' for the $\be\ga$-system.
The vacuum of the $q$ picture has $Q_{3/2} = q$.  For $\ka=\half$,
$q$ takes integer values, and for $\ka = 0$, half-integer.
We make a fixed choice for the vacuum of the
spin $\half$ fields; namely, for NS we use the $SL(2,\RR)$ vacuum, and
for R we demand $\psi^\pm_n\,(n>0)$ and $\psi^-_0$ annihilate the vacuum.
Similarly, we choose once and for all the $bc$ ghost
system's vacuum to be $c_1|0\rangle$, where $|0\rangle$ denotes
its $SL(2,\RR)$ vacuum.  Neither of these fixed choices is
displayed in the notation.  In this way,
the ``physical vacuum" $|p^M,ip^L; -\ka - \half\rangle$ ($\ka = 0,\half$)
used in the cohomology calculation
of Section 2,  is exactly that for which all negatively moded oscillators are
creation operators.  The ``total" ghost number ${\rm gh}
\equiv Q_{3/2} + Q_{2}$ is normalized so that $d$ has ghost number one,
and the physical vacuum has ghost number zero (independent of the sector).
\foot{Note that, a priori, choosing the NS vacuum to have ghost number zero
would fix the R vacuum to have ghost number $-\half$. To have a more symmetric
result for the cohomology we have decided to normalize the ghost number
operator differently in each sector.}
\ss

After bosonization of the half-integer-spin fields, which
is discussed in Appendix C, the vacuum states
are labeled by the charges $Q_{\la}$.  The spin 1 fermionic ghost
system which arises is by definition always in the $SL(2,\RR)$ vacuum
sector.\ss

For the computation of cohomology it is convenient to introduce a
set of ``lightcone'' combinations of modes
\eqn\LZj{
\eqalign{
q^\pm  & = \sqrt{\half} (q^M\pm iq^L) \,,\quad\quad
p^\pm  = \sqrt{\half} \left( (p^M-Q^M) \pm i (p^L-Q^L) \right)\,, \cr
\al_n^\pm & = \sqrt{\half} (\al_n^M \pm i\al_n^L ) \,,\quad n\neq0
\,,\quad\quad \psi_m^\pm = \sqrt{\half} (\psi_m^M \pm i\psi_m^L ) \,. \cr}
}
with nonvanishing commutation relations
\eqn\LZk{
[q^\pm,p^\mp] =i\,,\qquad [\al_m^\pm,\al_n^\mp ] = m\de_{m+n,0}\,,\qquad
\{\psi_m^\pm,\psi_n^\mp\} = \de_{m+n,0} \,,
}
as well as shifted momenta
\eqn\pbshift{
P^\pm(n) = \sqrt\half \left( (p^M-(n+1)Q^M) \pm i(p^L-(n+1)Q^L) \right) \,.
}
In particular $p^\pm =  P^\pm(0)$.

In terms of these the BRST operator $\whd$ is given by
\eqn\pbbrst{\eqalign{
\whd = & \sum_{n\neq0} c_{-n}\left( P^+(n)\al^-_n + P^-(n)\al^+_n\right)
+ \sum_{m\in\ZZ+\ka} \ga_{-m}\left( P^+(2m)\ps^-_m + P^-(2m)\ps^+_m\right)\cr
& +\sum_{m,n\in\ZZ;m,n,m+n\neq0}
c_{-n}\left( \al^+_{-m}\al^-_{m+n} +\half (m-n)c_{-m}b_{m+n} \right)\cr
& + \sum_{n\in\ZZ,m\in\ZZ+\ka; m,n,m+n\neq0}
c_{-n}\left(  (\half n+m)\ps^+_{-m}\ps^-_{m+n} + (\half n-m)\ga_{-m}
\be_{m+n} \right)\cr
& + \sum_{n\in\ZZ,m\in\ZZ+\ka; m,n,m+n\neq0}
\ga_{-m} \left( \al_{-n}^+\ps^-_{n+m} + \al_{-n}^-\ps^+_{n+m}
- b_{-n}\ga_{m+n} \right) \,. \cr}
}


\appendix {B} {Cohomology computations}

This appendix  consists of three parts. First we present a summary of
mathematical results on the computation of cohomology using a spectral
sequence technique. Their proofs can be found in standard textbooks on
homological algebra (\eg [\BT,\HS]),  but  are usually presented at a more
abstract level than the pedestrian approach adopted here.
In the simplest cases, in particular those pertinent to the
2D (super) string, elementary proofs of some of
the results below have been discussed
recently in [\BMPd,\BMPe] and [\IO]. The second part reviews these
computations of the relative cohomology on $\fff$,
as a concrete application of these general techniques.  Finally, in the
third part, we discuss the calculation of the absolute cohomology.

\medskip\noindent
{\it B.1. Cohomology of a filtered (graded)  complex}
\medskip

Consider a complex  $(\cC,d)$ of vector spaces, where
$\cC=\bigoplus_n\cC^{(n)}$ and the differential $d:\lC{n} \to\lC{n+1}$.
We will consider a spectral sequence which arises when there is
an additional gradation, such that for each order $n$,\foot{This
is stronger than the usual assumption that $\cC$ must be a filtered
complex. A standard filtration in our case is given by subspaces
$\cC_{(k)}=\bigcup_{k'\geq k} \cC_k$.}
\eqn\SSaa{\lC{n}=\bigoplus_{k\in \ZZ} \lC{n}_k \,.} We will refer to the
integer $k$ as the degree, and denote the projection onto the
subspace of degree $k$ by $\pi_k$. This gradation by the degree must
satisfy the following properties:\smallskip

\item{1.} The differential $d$ has only terms of nonnegative degree,
\ie \eqn\SSab{d=d_0+d_1+\ldots=d_0+d_>\,,} where
\eqn\SSac{d_i:\lC{n}_k\to\lC{n+1}_{k+i}\,.} \item{2.} In each order
only a finite number of nontrivial degrees are present, \ie for each
$n$, spaces $\lC{n}_k$ are nontrivial for a finite number of $k$'s.

The problem is to set up a systematic method of computing cohomology
classes of $d$, which we denote by $H_d(\cC)$. The first observation is
that $d^2=0$ implies \eqn\SSnil{\summ{i,j}{i+j=k}d_id_j=0\,,\quad
k=0,1,\ldots\,} and, in particular,
\eqn\SSnzer{
d_0^{\,2}=0\,.} Thus we can
consider another complex $(\cC,d_0)$, with the same underlying space
$\cC$ and $d_0$ as the differential. Note that $(\cC,d_0)$ is in fact a
direct sum of complexes labeled by the degree, and therefore its
cohomology is much easier to investigate. Moreover, there is a simple
necessary condition for the cohomology of $d$ being nontrivial,
namely  $H^{(n)}_d(\cC)=0$ whenever $H^{(n)}_{d_0}(\cC)=0$ (see [\BMPd]).

Any $\psi\in\cC^{(n)}$ can be expanded as finite sum of terms
with definite degree,  $\psi=\psi_k+\psi_{k+1}+\ldots + \psi_p$.
By examining the condition $d\psi=0$ in each degree it is
also easy to prove [\BMPd] that  one can always choose representative
$\psi=\psi_k+\ldots+\psi_p$ of a nontrivial cohomology class in
$H^{(n)}_d(\cC)$ such that the lowest degree term $\psi_k$ in $\psi$
represents a nontrivial cohomology class in $H^{(n)}_{d_0}(\cC)$.

A spectral sequence is a ``gadget'' that allows a systematic investigation
of which $\psi_k\in H^{(n)}_{d_0}(\cC)$ extend to a cohomology class
$\psi\in H^{(n)}_{d}(\cC)$, where $\psi=\psi_k+\psi_>$ and $\psi_>$
stands for  ``corrections'' of degree higher than $k$.  The $r$th term
in the spectral sequence is simply the space of those $\psi_k + \cdots
+ \psi_{k+r}$ which survive this extension ``nontrivially" through $r$
degrees.  The successive terms then give finer and finer approximations
to $H^{(n)}_{d}(\cC)$.  We will
first establish the construction of successive terms of the
spectral sequence, but the reader will recognize where the
definitions are heading by keeping in mind the obvious
necessary conditions at each degree imposed by $d\psi = 0$.

The first term of the spectral sequence associated with our
gradation is
$ E_1=H_{d_0}(\cC)$. Next observe that   \SSnil\ for $k=1,2$
gives \eqn\SSntw{\eqalign{ d_0d_1+d_1d_0 &=0\,,\cr
d_0d_2+d_1d_1+d_2d_0 &=0\,.\cr}} The first equation tells us that $d_1$
induces a well defined transformation $d_1:E_1\to E_1$, while the
second says that this induced map is nilpotent, $d_1^{\,2}=0$, on $E_1$. Call
$\de_1=d_1$ and consider the complex $(E_1,\de_1)$.

The second term, $E_2$,  in the spectral sequence is equal to the
cohomology of the previous term, namely
$E_2=H_{\de_1}(E_1)$.  This one also has a canonical
differential obtained from $d$, which is explicitly constructed as
follows:
\hb
Let $\psi_k\in \cC$ represent an element of $E_2$. Then  we must have
\eqn\SSrep{\eqalign{d_0\psi_k&=0\,,\cr
d_1\psi_k+d_0\ps_{k+1}&=0\,,\cr}} for some $\ps_{k+1}\in\cC_{k+1}$.
The first condition is needed so that $\psi_k$ defines an element in
$E_1$, while the second is simply $\de_1\psi_k=0$ in $E_1$.  Define
$\de_2:E_2\to E_2$ by
\eqn\SSdetw{\eqalign{\de_2\psi_k&=d_2\psi_k+d_1\ps_{k+1}\cr
&=\pi_{k+2}d (\psi_k+\ps_{k+1})\,.\cr}} With a little of work one
verifies that $\de_2$ does not depend on any of the choices made, and
that $\de_2\de_2=0$. Thus we can define $E_3=H_{\de_2}(E_2)$, and so
on.

In general,  $\psi_k\in \cC$ represents an element in $E_r$, $r>1$, if
there exist $\ps_{k+1},\ldots,\ps_{k+r-1}$ such that \eqn\SSrter{\pi_i
d(\psi_k+\ps_{k+1}+\ldots+\ps_{k+r-1})=0\,,\qquad {\rm for} \quad
i=k,\ldots,k+r-1\,.} Then $\de_r:E_r\to E_r$,
\eqn\SSdifr{\de_r\psi_k=\pi_{k+r}
d(\psi_k+\ps_{k+1}+\ldots+\ps_{k+r-1})\,,} is well defined and
nilpotent. We now see precisely what it means that
the elements of $E_r$ are those cohomology
classes of $d_0$ which can be extended to approximate cohomology
classes of $d$ through $r$ degrees.

The spectral sequence in this context is simply the sequence of
complexes $(E_r,\de_r)_{r=1}^\infty$ constructed as above. In
interesting cases this sequence converges, which means that the spaces
$E_r$ stabilize, \ie \eqn\SSstab{E_r=E_{r+1}=\ldots =E_\infty\,,} for
some $r\geq 1$.  Obviously, this requires
\eqn\SSdst{\de_r=\de_{r+1}=\ldots=\de_\infty=0\,.} In such a case one
also says that the sequence collapses at $E_r$.

Thus, as stated, computing subsequent terms in the spectral sequence is
nothing other than systematically correcting $\psi_k\in H_{d_0}(\cC)$
such that, if it lives till $E_\infty$, $\psi_k+\psi_>$  should
represent a class in $H_{d}(\cC)$.  Indeed, the main theorem in the subject is
that under the assumptions above [\BT,\HS]
\eqn\KPlimit{
E_\infty\simeq H_d(\cC)\,.
}
For example, \SSrep\ is precisely the statement that the nontrivial
$d_0$-cohomology state $\psi_k$ may be corrected by a degree $k+1$
term so that the result is annihilated by $d$ through degree one terms.
But moreover, the second term of the spectral sequence also throws
away states which are $d_1$ trivial in $d_0$ cohomology, which
is equivalent to demanding the extension is nontrivial to this order.
Similar arguments may be given at higher order.

It is useful to keep the following two observations in mind.\smallskip

\noindent 1. One might get the wrong impression that the magic of the
spectral sequence absolves one from doing any work. This is not true.
At each stage one must compute the cohomology of $\de_r$ on $E_r$, and
this may become quite complicated.  Rather, the spectral sequence
provides an algorithm for a systematic computation, which can be
formulated quite abstractly and then applied in different settings.

\noindent 2. Most spectacular applications of these techniques are in
cases when one needs to do very little calculations to get an answer.
In particular, a rather trivial observation is that $\de_r$ increases
the degree by $r$, so often one can deduce that it is zero by simply
examining the set of degrees in which $E_r$ is nontrivial. For example,
if all of the  cohomology $H_{d_0}(\cC)$ is concentrated in a single
degree  $k$ (note that this allows for various orders  $n$ with
nontrivial $H^{(n)}_{d_0}(\cC)$) then the sequence must collapse at the
first term, \ie $E_1\simeq E_\infty$, or, equivalently,
$H^{(n)}_{d_0}(\cC)\simeq H^{(n)}_{d}(\cC)$.

\bigskip
\noindent
{\it B.2. The relative cohomology of $\fff$}
\medskip

A simple application of the general formalism described above is the
computation of the relative BRST cohomology in Section 2.
We will consider NS and R sectors separately.

In the NS sector the calculation initially follows that in [\BMPd] for
the bosonic case.  The lightcone combinations \LZj\ allow us to assign
a degree to the oscillators,

\eqn\PBi{\eqalign{
{\rm deg} (\al_n^+)&= {\rm deg} (c_n)= {\rm deg} (\ps_r^+)= {\rm deg}
(\ga_r)=+1\,,\cr
{\rm deg} (\al_n^-)&= {\rm deg} (b_n)= {\rm deg} (\ps_r^-)= {\rm deg}
(\be_r)=-1\,,\cr}}
under which $\whd$ decomposes as
\eqn\JOb{
\whd = \whd_0 + \whd_1 + \whd_2 \,.}
Here $\whd_k$ denotes terms with degree k, and, in particular,
\eqn\PBj{
\whd_0 = \summ{n\in\ZZ}{n\neq0} P^+(n) c_{-n}\al_n^- + \sum_{m\in\ZZ+\half}
P^+(2m) \ga_{-m}\ps^-_m \,.}
The coefficients $P^\pm(n)$ are given by \pbshift.
Note that the parametrization \JOa\ follows from the
condition $P^+(r) = P^-(s) = 0$ for integers $r$ and $s$.

Consider now the proof of the results of Section 2.  The exceptional
case (i) has been discussed in Section 2.  When $P^+(n) \neq 0 \quad
\forall n \neq 0$, a ``contracting homotopy" can be constructed,
and case (ii) follows immediately.  In fact, nontrivial exceptions
can only arise when there are a pair of integers $r$ and $s$, $rs > 0$,
such that both $P^+(r)=P^-(s)=0$.  [For more details see [\BMPd] for the
directly analogous discussion in the bosonic case.]  The calculation
now splits into several cases, whose proofs all run quite parallel to each
other.  Thus we consider first the case $r,s \in \ZZ_+$, \ie case (iii),
and then we
must further distinguish between even and odd values.

For $r \in 2\ZZ_+$, the $\whd_0$ cohomology is built
from the oscillators $\al_{-r}^+$ and $c_{-r}$, and at the level
${rs\over2}$ ($L_0=0$) there are only two such states (both of the same
degree) for $s \in 2\ZZ_+$, but none otherwise.  Clearly in this
case the cohomology of $\whd_0$ coincides with that of $\whd$
(see remark 2 below \KPlimit).

For $r \in 2\ZZ_+ - 1$, the cohomology  of $\whd_0$
is spanned by the states of the form
\eqn\dupa{
(\al_{-r}^+)^{a_1}(\psi_{-{r\over 2}}^+)^{a_2}(c_{-r})^{b_1}
(\ga_{-{r\over 2}})^{b_2} |p^M,ip^L; - \half\rangle \,,}
where the integers $a_1$, $a_2$, $b_1$ and $b_2$ satisfy
\eqn\apud{
(a_1+b_1) +\half(a_2+b_2)={s\over 2}\,,\quad a_1,b_2\geq 0\,,\quad
a_2,b_1=0,1\,.}
Thus the cohomology of $\whd_0$ clearly is {\it not}
generally restricted to a single degree, and further analysis is required.
We identify the space spanned by the states in
\dupa\ with the first term $E_1$ of the spectral sequence.
The induced differential $\de_1$ is explicitly given by
\eqn\dupnydiff{ \de_1=\psi^+_{-{r\over 2}}\ga_{-{r\over 2}}\al_r^- -
\ga_{-{r\over 2}}\ga_{-{r\over 2}}b_r\,.}
By examining the action of $\de_1$ on the states \dupa,
one finds that if $s \in 2\ZZ_+ - 1$
the cohomology of $\de_1$ on $E_1$ is concentrated
in only one degree, and is two dimensional with one state at ghost
number 0 and 1, respectively.  Otherwise the cohomology of $\de_1$ on $E_1$
vanishes, and thus no nontrivial state is obtained.
[An elementary derivation of these results,
which in fact is equivalent to the above spectral sequence argument, has
recently been given in [\IO].]  This completes the analysis for case
(iii), and case (iv) is analogous.

For the R sector there is one essential difficulty, beyond the NS
calculation above, due to the fact that $G_0$ does not act reducibly on
$\fff$.  Thus the restriction to the
$\cF_\rel(p^M,p^L)$ subspace is not trivial
to carry out.  Of course in the case that either $r$ or $s$ vanishes, the
$\whd$ cohomology on the whole Fock spaces is anyway at most
one vacuum state (by the same arguments as in [\BMPd]) and case (ii)
again follows easily. The other cases may be dealt with by introducing
``rotated" oscillators  in terms of which  $\cF_\rel$ can be
constructed explicitly.

The operator $\vartheta = \psi_0^+/p^+$ is well defined for
$r\not=0$, and satisfies \eqn\JOd{\{G_0,\vartheta\} = 1 \,.}
Of course $G_0^{\,2}=0$ on $\cF^{L_0}(p^M,p^L)$,
so the existence of such an operator implies
that the cohomology of $G_0$ on $\cF^{L_0}(p^M,p^L)$
is trivial, and indeed $\vartheta$ is
the corresponding contracting homotopy. In particular we have
\eqn\kutas{\cF_\rel=G_0\cF^{L_0}\,,\quad \cF^{L_0}=\cF_\rel\oplus
\vartheta\cF_\rel\,,\quad {\rm dim}\,(\cF_\rel)=
 {\rm dim}\,(\vartheta\cF_\rel)= \half {\rm dim}\,(\cF^{L_0})\,.} Further,
if $o_m$ denotes any fundamental oscillator, we may define rotated
oscillators [\IMNU] \eqn\JOe{\widetilde{o}_m \equiv
[G_0,\vartheta o_m\} \,,}
which again satisfy the same algebra as the original oscillators
$o_m$ (except that $\widetilde \psi_0^+ = 0$ by this definition).

 If ${\cal O}$ is an operator built from the fundamental
oscillators, normal ordered with respect to the physical vacuum
with zero modes of $\psi^\pm(z)$ to the right,
we denote by $\widetilde{\cal O}$ the operator obtained by
rotating all the oscillators in $\cal O$. This is achieved by
$\widetilde{\cal O}= [G_0,\vartheta {\cal O}\}$, as can easily be seen
by pulling the commutator through all the oscillators; for example,
if we have two even oscillators $o_1$ and $o_2$,
$\{G_0, \vartheta o_1o_2\} = o_1o_2 - \vartheta [G_0,o_1]o_2
-\vartheta o_1[G_0,o_2] = \widetilde{o}_1\widetilde{o}_2$ upon
using $\vartheta^2 = 0$.  Since
$\widetilde{\cal O}={\cal O}-\vartheta[G_0, {\cal O}\}$, we find that
any operator commuting with $G_0$ can be rewritten in terms of the
rotated oscillators.  In particular  this applies to $\whd$.

Moreover, we may also show that $\cF_\rel$ is contained in
the subspace freely generated by the rotated
creation oscillators  acting on the vacuum $\state{p^M,ip^L,-\half}$. Indeed,
let $\psi={\cal O}\state{p^M,ip^L,-\half}$ denote a state in this
subspace, where ${\cal O}$ is some  operator built with creation
operators with respect to this vacuum. Then
\eqn\hujek{
\psi=G_0\vartheta{\cal O}\state{p^M,ip^L;-\half}
=\widetilde{\cal O}\state{p^M,ip^L,-\half}\pm \vartheta{\cal O}
G_0\state{p^M,ip^L;-\half}\,,
}
and by examining the zero modes
$\psi^\pm_0$ in $G_0$ one easily verifies that the last term vanishes.

This analysis applies to all states built on the Fock space with
$p^+ \neq 0$, \ie $r \neq 0$.  When $r =0$, but $s \neq 0$ we may
go through precisely the same analysis, but using
$\bar{\vartheta} = {{\psi_0^-}\over{p^-}}$, with the obvious changes.
Thus one effectively needs to use two ``patches" in this oscillator space.
The exceptional case $p^+=p^-=0$ has been dealt with separately in Section 2.

With these results in hand, the computation of cohomology
is straightforward. We introduce a filtration on $\cF_\rel$ by assigning
degrees
as in \PBi\ but to the  rotated oscillators.
Furthermore, since $\widetilde{\psi}^+_0 = 0$, only
non-zero modes appear in $\whd_0$,
\eqn\JOg{
\whd_0 = \sum_{n \neq 0} \big[\,P^+(n)\widetilde{c}_{-n}\widetilde{\al}^-_n
+ P^+(2n)\widetilde{\ga}_{-n}\widetilde{\psi}^-_n \,\big]\,.}
The rest of the computation proceeds as in NS sector, in particular
one must consider at most the second term of the spectral sequence.

\bigskip
\noindent
{\it B.3.  The absolute cohomology of $\fff$}

In this subsection we give the remaining details for the proof of the
relation between the relative and absolute cohomologies in R sector
discussed in Section 2.  Our approach has been inspired by  that of [\Hx]
for the case of ten dimensional superstring.

First note
that any state $\varphi$ in the product of Fock
spaces $\fff$ corresponding to the $q=-\half$ picture may be decomposed with
respect to ghost zero modes $c_0$ and $\ga_0$ as
\eqn\KPabsolut{\varphi = \phi + c_0\psi\,, \quad b_0\phi = b_0 \psi =0 \,,}
and
\eqn\KPblabla{\eqalign{
\ph & = \sum_{n=0}^{n(\ph)} \ga_0^n \phi_n \,,\quad \be_0\phi_n = 0\,,\cr
\psi & = \sum_{n=0}^{n(\ps)} \ga_0^n \psi_n \,,\quad \be_0\psi_n = 0\,,\cr}}
where $n(\ph)$ and $n(\ps)$ are finite. [Strictly speaking the
finiteness of  the expansion into series in $\ga_0$  should be viewed as
a definition of the Fock space topology.] We will denote
$(\psi)_>=\psi-\psi_0$.

Let us consider the case in which one of $p^+$ or $p^-$
is nonzero.  As we saw in the previous section,
the cohomology of $F=G_0$ on $\cF^{L_0}(p^M,p^L)$ is then trivial.
This allows us to prove the following technical result about  the
states in ${\cal K}(p^M,p^L)$:

For $\psi\in {\cal K}(p^M,p^L)$, \ie
$\psi = \sum_{m=0}^{n(\psi)} \ga_0^m \psi_m \,,\quad \psi_m \in
\cF^{L_0}(p^M,p^L)$,
a general solution to \eqn\dbareq{(\bd\psi)_>=0 \,,}
is of the form \eqn\dbarres{\psi=\bd \rho + F\rho'_0\,,}
where $ \rho,\rho'_0 \in {\cal K}(p^M,p^L)$ and in addition
$\be_0\rho'_0=0$ and $n(\rho)= n(\psi)-1$.

\noindent
In fact, recall that $\bd=\ga_0 F + \be_0 N +\whd$, so that $\bd\psi=0$
written in components reads ($n=n(\psi)$)
\eqn\system{\eqalign{
\whd\psi_0-N\psi_1&=0\cr
\whd\psi_1+F\psi_0-2N\psi_2&=0\cr
&\vdots\cr
\whd\psi_{n-1}+F\psi_{n-2}-nN\psi_n&=0\cr
\whd\psi_n+F\psi_{n-1}&=0\cr
F\psi_n&=0\,.\cr}}
We must show that solving  all, but the first, of the above equations
yields $\psi $ of the form  \dbarres. Clearly the
solution to the last equation is $\psi_n=F\chi_{n}$. Since
$\{F,\whd \,\}=0$, the next equation gives
$F(\whd\chi_{n}-\psi_{n-1})=0$, which is solved by  $\psi_{n-1}=F\chi_{n-1}
+\whd\chi_{n}$. Continuing, we obtain
\eqn\KPsol{
\psi_k=F\ch_k-(k+1)N\ch_{k+2}+\whd\ch_{k+1}\,,\quad k\geq
0\,, \quad \chi_i=0\,,i>n\,.
}
Introducing
\eqn\KPform{
\rho=\sum_{m=0}^{n-1} (\ga_0)^m\chi_{m+1}\,,\qquad
\rho_0=\ch_0\,,
}
we obtain \dbarres.

Using this result it is easy to verify that any absolute cohomology
state $\varphi=\ph+c_0\psi$  is equivalent to a linear combination
of states  \JRe\ given by the relative cohomology.
Indeed, in components, $d\varphi=0$ reads
\eqn\KPzero{\bd\phi-(M+\ga_0\ga_0)\psi=0\,,\quad
\bd \psi=0\,,}
while the ``gauge transformation''
$\varphi\rightarrow\varphi+d\omega$,  where $\omega =\xi+c_0\chi$, is
\eqn\KPgauge{
\phi\rightarrow \phi+\bd \xi-(M+\ga_0\ga_0)\chi\,,\quad
\psi\rightarrow \psi-\bd\chi\,.
}
Since $\bd\psi=0$,  by setting $\ch=\rho$ one can immediately
gauge away all terms  in $\psi$ except $\psi_0$. Thus we may assume
$\psi=\psi_0$, where $\whd\psi_0 = F\psi_0 =0$. The remaining
gauge freedom is
\eqn\KPremgauge{
\psi_0\rightarrow \psi_0+\bd\chi \,, \quad\quad (\bd\ch)_>=0 \,.
}
Thus $\chi$ must be of the form $\ch=\bd\sigma+F\sigma'_0$, and
\KPremgauge\ simply
becomes $\psi_0\rightarrow  \psi_0+\whd F\sigma'_0$; \ie
only the relative cohomology component in $\psi_0$ cannot be gauged away.
{}From the discussion in Section 2, we know that if $\psi_0$ is a
relative cohomology state, there exists $\la=\la_0+\ga_0\la_1$
such that $d(\la+c_0\psi_0)=0$.
Writing $\varphi=(\phi-\la) + (\la+c_0\psi)$,
there is an $\omega$ as above such that
$\varphi=(\phi'-\la) + (\la+c_0\psi_0) + d\omega$, and
we deduce  from \KPzero\ that $\bd(\phi-\la)=0$. Therefore all terms
in $\phi-\la$, with the possible exception of the lowest order  one, can be
gauged away using a suitable $\xi$ in \KPgauge. This concludes the proof
that any  absolute cohomology class can be written as a sum of two
($d$ closed states) of the form \JRe\ , canonically constructed from relative
cohomology classes.

We must still verify that  for a given relative cohomology
representative  $\psi_0$, the corresponding  states, $\psi^{(1)}$ and
 $\psi^ {(2)}$ defined in \JRe\ are nontrivial representatives of the
absolute cohomology.  For $\psi^{(2)}$ this clearly follows from the
discussion above.  In the case of  $\psi^{(1)}$, we certainly cannot
gauge it away using $\xi$ alone. Thus we must consider the most general
gauge transformation.  Remarkably,  using the general solution for
$\ch$ (which clearly must satisfy $\bd \chi=0$ so that no ``$c_0$"
term is induced by the gauging), and the identities \commutatb, it is
straightforward to verify  that $(M+\ga_0\ga_0) \chi= \bd \ch'$, for some
$\ch'$. [In particular, using \dbarres\ and further counting of available
ghost numbers for relative cohomology, $\bd \chi = 0$ implies
$\chi = \bd \sigma$ here.]  This shows that $\psi^{(1)}$ is also nontrivial.
\bs


\appendix {C} {Bosonization}

Our cocycles for the bosonized operators, and thus the spin fields,
are constructed as follows.  The basic rule is that all
``fermionic" first order fields anticommute (in this regard we follow
the conventional choice of [\KLLSW]).  A ``minimal" set
may be obtained explicitly as
\eqn\JAl{
\psi^\pm(z)=e^{\pm iH} I^{\pm 1} \,,\quad
\ga(z)= e^{i\tilde \ph} \tI \eta \,,\quad
\be(z)=\p\xi e^{-i\tilde \ph} \tI^{-1}\,,
}
where
\eqn\JAm{
I = e^{i\pi(Q_1 + Q_2)} \,,\quad \tI = e^{i\pi(Q_1 + Q_2 + Q_{1/2})} \,.
}
Here $Q_n$ just denotes the fermion number corresponding to a spin $n$ pair.
The spin fields are given the ``obvious" square root cocycle,
\eg $S^{\pm} = e^{\pm iH/2} I^{\pm {1\over2}}$.
Note that writing out the cocycle operators is not necessary in
calculations, but useful for setting consistent definitions.
With the above we find, for example
\eqn\JAn{
\tI^{-m} e^{inH(z)} \tI^m = e^{-imn\pi} e^{inH(z)} \,.
}
For the calculations in Section 3, however, we found it useful to enlarge
the set of mutually anticommuting operators to include $e^{\pm i \phi^M}$
and $e^{\pm \phi^L}$ -- in particular this allows a choice so that the
$sl(2)$ symmetry currents \JTj\ are ``manifestly bosonic".
Moreover, for Section 4 it is necessary to include
further cocycles so that the first order fields mutually anticommute
between left and right moving sectors.  The complete set of cocycles used
for these calculations are defined by \JAl\ together with
\eqn\JAnn{
\eqalign{
e^{\pm i\phi^M} & \ara e^{\pm i\phi^M}I_M^{\pm 1} \,,\cr
e^{\pm \phi^L} & \ara e^{\pm \phi^L}I_L^{\pm 1} \,,\cr
}
}
and similar definitions for the right moving sector distinguished
by $'$, where
\eqn\JAno{
\eqalign{
I &= \exp\{i\pi(Q_1+Q_2-Q_1{}'-Q_2{}'-Q_{{1\over2}}{}'-Q_{{3\over2}}{}'
-p^M{}' -ip^L{}')\} \cr
\tI &= \exp\{i\pi(Q_1+Q_2+Q_{{1\over2}}-Q_1{}'-Q_2{}'-Q_{{1\over2}}{}'-
Q_{{3\over2}}{}' -p^M{}' -ip^L{}')\} \cr
I' &= \exp\{-i\pi(Q_1{}'+Q_2{}'-Q_1-Q_2)\} \cr
\tI' &= \exp\{-i\pi(Q_1{}'+Q_2{}'+Q_{{1\over2}}{}'-Q_1-Q_2)\} \cr
I_M &= \exp\{i\pi (Q_1+Q_2+Q_{{1\over2}}+Q_{{3\over2}}-Q_1{}'-Q_2{}'-
Q_{{1\over2}}{}'-Q_{{3\over2}}{}'-p^M{}'-ip^L{}')\} \cr
I_M' &= \exp\{-i\pi (Q_1{}'+Q_2{}'+Q_{{1\over2}}{}'+Q_{{3\over2}}{}'
-Q_1-Q_2)\} \cr
I_L &= \exp\{i\pi (Q_1+Q_2+Q_{{1\over2}}+Q_{{3\over2}} +p^M -Q_1{}'-Q_2{}'-
Q_{{1\over2}}{}'-Q_{{3\over2}}{}'-p^M{}'-ip^L{}')\} \cr
I_L' &= \exp\{-i\pi (Q_1{}'+Q_2{}'+Q_{{1\over2}}{}'+Q_{{3\over2}}{}'
+p^M{}'-Q_1-Q_2)\}
\,. \cr
}
}
As an aside, note that if suitable cocycles were included in the
minimal set \JAm\ to ensure left and right mutual anticommutation of the
appropriate fields, then the operators $a_1$ and $a_2$ of Section
4 -- which generate the ground ring of the uncompactified matter model --
would still be mutually commuting.  However, the remaining
generators -- as required for the compactified matter model --
commute up to relative phases, making the determination of
the symmetries induced for the uncompactified model more
difficult.\bs

Recall that since $\ga$ and $\be$ commute with the operator
$Q_P \equiv Q_{3/2} + Q_1$, $Q_P$ has eigenvalue $q$ for every
$\be\ga$ Fock space state in
a given $q$ picture.  Thus, after bosonization, the $q$ picture
is just found as that ``slice" of the complete bosonized space with a definite
eigenvalue $q$ for $Q_P$.
The picture changing operator is
\eqn\JMb{
X(z) = \{d,\xi(z)\} \,,
}
which has $Q_{3/2} + Q_2 = 1$,  and $Q_1 - Q_2 = 0$.
It clearly increases the picture by one unit, commutes
with $d$, and has $L_0 = 0$.
The bosonized expression for $X(z)$ is
\eqn\JAc{
X(z) = (G^M+G^L)e^{i\tphi} + c\p\xi -be^{2i\tphi}\p\eta -
\p(be^{2i\tphi}\eta) \,.
}
The $d$-invariant operator with $L_0 = 0$ which decreases the
picture by one unit is denoted $Y(z)$
\eqn\JMz{
Y(z) = c\p\xi e^{-2i\tphi} \,.
}
It has $Q_{3/2} + Q_2 = -1$ and $Q_1 - Q_2 = 0$.

The explicit action on the states was clearly discussed in [\HMM].
The zero modes $X_0$ and $Y_0$ (\eg $X_0 = \oint {dz\over{2\pi iz}}X(z)$)
commute with each other.  More importantly
\eqn\JMc{
:X_0::Y_0: = 1 + [d,\epsilon] \,,
}
for some operator $\epsilon$, and thus they are inverse in cohomology.
Similarly
\eqn\JMd{
:X_0:^n:Y_0:^n = 1 + [d,\epsilon'] \,,
}
and we can go from one picture to another with these operators
which clearly provide an isomorphism in cohomology.


\appendix {D} {Explicit representatives for $\hat{c}=1$}

In this appendix we will present some results on the structure of the
$\hat{c}=1$ Fock space modules of the $N=1$ superconformal algebra.
We will show how to obtain expressions for the singular vectors in terms
of certain ``super-Schur polynomials,'' and how these can be used to obtain
explicit representatives of the BRST-cohomology. The line of reasoning closely
follows the discussion in the bosonic case [\BMPe].

{}From the Kac determinant [\Ka] and the existence of a hermitian form, it
follows that the $\hat{c}=1$ Fock space $\cF(p)$ is reducible if and only if
$p= \half (r-s)$ for some $r,s\in\ZZ\,,rs>0\,,(r-s)\in2\ZZ+ (1-2\ka)$.
In this case the Fock space contains a singular vector at level ${rs\over2}$.
Moreover, in this case, the Fock space module is completely reducible, \ie
$\cF(p=\half(r-s)) = \bigoplus_{\ell\geq0} {\cal L}(h=
{1\over8} (|r-s|+2\ell)^2,\hat{c}=1)$.

The submodules obtained by restricting this sum to $\ell\geq\ell_0\geq0$
are isomorphic to $\cF(\half(r_k-s_k))$, where $k=\max(0,s-r) + \ell_0\,,
r_k=r+k\,,s_k=s-k$. The isomorphism is given by the operator
$(J_-)^k$, where
\eqn\pbapa{
J_- ={1\over 2\pi i} \oint dz\ \ps(z) e^{-i\ph(z)}\,,
}
is the ``screening charge'' for the $\hat{c}=1$ free field realization.
As we have seen, $J_-$ also equals the negative root operator of an
$sl(2)$ algebra. Clearly, this construction of singular vectors is
intimately related to the fact that
the $N=1$ superconformal algebra is the commutant of the horizontal
$sl(2)$ algebra on $c={3\over2}$ highest weight modules of affine $sl(2)$.
The isomorphism $(J_-)^k$ can be used to obtain explicit formulas
for the singular vectors. Let us first introduce some special functions.

Elementary Schur polynomials $S_k(x)\,,x=(x_1,x_2,\ldots)\,,k\geq0$ are
defined through a generating function
\eqn\pbapb{
\sum_{k\geq0} S_k(x) z^k = \exp \left( \sum_{k\geq1} x_kz^k \right)\,.
}
For convenience we put $S_k(x)=0$ for $k<0$.
They can be generalized to the supercase by introducing Grassmannian
variables $\th_m$, where $m={1\over2},{3\over2},{5\over2},\ldots$ in the
NS-case and $m=1,2,\ldots$ in the R-case as well as an additional
Clifford variable $\th_0\,,\th_0{}^2=1$ in the R-case.

We now define NSR-Schur polynomials by
\eqn\pbapc{\eqalign{
S^{NS}_k(x,\th) & = \sum_{l\geq {1\over2}} S_{k-l}(x) \th_l\,, \cr
S^R_k (x,\th) & = \sum_{l\geq0} S_{k-l}(x)\th_l\,,\cr}
}
in terms of which we can expand
\eqn\pbapd{
\ps_<^{NS/R}(z) e^{-i\ph_{<}(z)}= -\sqrt\half \sum_{m\geq\ka}
S_m^{NS/R}(x,\th) z^{m-{1\over2}}\,,
}
where
\eqn\pbape{
\ps_<^{NS/R}(z) = \sum_{m\leq -\ka} \ps_m^{NS/R} z^{-m-{1\over2}}\,,\qquad
\ph_<(z) = \sum_{n<0} {1\over n} \al_n z^{-n} \,,
}
and $x_j = -\al_{-j}/j\,,\th_m = -\sqrt2 \ps_{-m}$.

Now suppose $r,s\in\ZZ_+$.
As we have argued above, the singular vector $\Ps_{(r,s)}$
at level ${rs\over2}$ in
$\cF(p=\half(r-s))$ is given by \foot{For R, one can take either of the
two possible vacua $\state{\pm}$ where $\state{+} = \ps_0 \state{-}$.}
\eqn\pbapf{
\left( {1\over2\pi i}\oint dz\
\ps(z) e^{-i\ph(z)} \right)^s |\half(r+s)\rangle\,.
}
After normal ordering and expanding the resulting expression by making
use of \pbapd, the contour integrals can be evaluated. This results in
explicit expressions for the singular vectors in terms of the NSR-Schur
polynomials \pbapc. For example
\eqn\pbapg{
\eqalign{
\ps_{(r,1)}^{NS} & = S_{{r\over2}}^{NS}(x,\th) |\half(r-1)\rangle
\,,\qquad r\in2\ZZ+1\,,\cr
\ps_{(r,2)}^{NS} & = \left( S_{{r\over2}}(x)S_{{r\over2}}(x)
+ S_{{r-1\over2}}^{NS}(x,\th)S^{NS}_{{r+1\over2}}(x,\th) \right) |\half(r-2)
\rangle \,,\qquad r\in2\ZZ\,,\cr
\ps_{(r,1)}^{R} & = S^R_{{r\over2}}(x,\th) | \half(r-1)\rangle
\,,\qquad r\in2\ZZ\,,\cr
\ps_{(r,2)}^R & = S^R_{{r-1\over2}}(x,\th)S_{{r+1\over2}}^R(x,\th)
|\half(r-2)\rangle\,, \qquad r\in2\ZZ+1\,. \cr}
}
In the NS case it is relatively
straightforward  to write expressions for generic $(r,s)$ in
terms of determinants of the elementary NS-Schur polynomials, as in the
bosonic case. In the R-sector there does not seem to be a comparably
simple expression, due to the fact that the R-Schur polynomials are not
mutually anticommuting because of the presence of the Clifford variable
$\th_0$. We refrain from giving more general expressions since the above
will suffice for the purpose of this paper.
The case $r,s\in\ZZ_-$ can be treated analogously, or simply by observing
that $\ps_{(-r,-s)}\sim \ps_{(s,r)}$. We also remark that,
because of a symmetry of $G_r$, the expression
for $\ps_{(s,r)}$ can be obtained from $\ps_{(r,s)}$ by letting
$(x,\th)\to (-x,-\th)$.
\bs

We will now discuss -- in complete analogy with the bosonic case --
how the above singular vectors immediately lead to representatives
of the (discrete) BRST cohomology at ghost number zero.

First of all, observe that
\eqn\pbaph{\eqalign{
[ d,\ps^Me^{\pm i\ph^M} ] &= \p \left(c\ps^Me^{\pm i\ph^M} \pm
\ga e^{\pm i\ph^M} \right)\,,\cr
[d , i\p\ph^M ] & = \p \left( c i\p\ph^M + \ps^M\ga \right) \,,\cr}
}
implying that the $sl(2)$ algebra generated by $\{ J_\pm, J_0\}$ (see
\JTj) acts on the cohomology. In particular, the BRST operator commutes
with the operator $(J_-)^k$ above, which provided the embedding
$\cF^M(\half(r_k-s_k)) \to \cF^M(\half (r-s))$. Moreover, since the
BRST operator $d$ acts on each irreducible constituent of $\cF^M(p)$
separately, we might as well restrict the computation of physical
states to each $\cF^M(\half(r_k-s_k))\otimes \cF^L(\half(r+s+2))
\otimes \cF^G$ (notice we defined $r_k,s_k$ such that $r_k+s_k = r+s$).
Choosing $k=s$ for $r,s>0$ (or $k=-r$ for $r,s<0$) we find that the
vacuum $|p^M,ip^L;q\rangle =
|\half(r_k-s_k),\half(r_k+s_k+2); -\half-\ka\rangle$
provides a nontrivial cohomology state. Since this state has $p^-=0$, it
is also in $\Ker G_0$ (for R), \ie it represents a relative cohomology state.
Mapping with the operator $(J_-)^k$ provides the nontrivial (relative)
cohomology state
\eqn\pbapi{
\ps_{(r,s)}(x^M,\th^M) |\half(r-s),\half(r+s+2);-\half-\ka\rangle\,.
}
By comparison to the results of Section 2, we conclude that these exhaust the
nontrivial (relative) cohomology states at ghost number zero.
So we see that, as in the bosonic case, it is possible to choose
a ``material gauge,'' \ie a representative only containing matter
excitations, for the ghost number zero sector. In particular
it also follows that the associated currents \JTg\
can be chosen to be superconformal primaries.

\footatend\immediate\closeout\rfile\writestoppt
\baselineskip=14pt{\bigskip\noindent {\bf  References}}%
\bigskip{\frenchspacing%
\parindent=20pt\escapechar=` \input refs.tmp\vfill\eject}\nonfrenchspacing

\vfil\eject\end